\documentclass{aastex62}

\newcommand{\hi}{\ion{H}{1}}      
\newcommand{\sif}{\ion{Si}{4}}   \newcommand{\sit}{\ion{Si}{3}}
\newcommand{\cw}{\ion{C}{2}}     \newcommand{\siw}{\ion{Si}{2}} 
    \newcommand{\kms}{\,km\,s$^{-1}$}
\newcommand{\hw}{\ion{H}{2}}     \newcommand{\sqcm}{\,cm$^{-2}$}
\newcommand{\tm}{\tablenotemark} \newcommand{\tn}{\tablenotetext}
\newcommand{\smy}{\,M$_\odot$\,yr$^{-1}$}

\graphicspath{{./}{figures/}}
\received{August 7, 2019}
\revised{August 27, 2019}
\accepted{August 29, 2019}
\submitjournal{ApJ} 
\shorttitle{Galactic Gas Flow Rates}
\shortauthors{Fox et al.}

\begin{document}

\title{THE MASS INFLOW AND OUTFLOW RATES OF THE MILKY WAY}

\correspondingauthor{Andrew Fox}
\email{afox@stsci.edu}

\author[0000-0003-0724-4115]{Andrew J. Fox}
\affil{AURA for ESA, Space Telescope Science Institute, 3700 San Martin Drive, Baltimore, MD 21218}

\author[0000-0002-1188-1435]{Philipp Richter}
\affil{Institut f\"ur Physik und Astronomie, Universit\"at Potsdam, Haus 28, Karl-Liebknecht-Str. 24/25, D-14476, Potsdam, Germany}

\author[0000-0002-6541-869X]{Trisha Ashley}
\affil{Space Telescope Science Institute, 3700 San Martin Drive, Baltimore, MD 21218}

\author[0000-0003-1127-7497]{Timothy M. Heckman}
\affil{Department of Physics and Astronomy, Johns Hopkins University, 3400 N. Charles Street, Baltimore, MD 21218}

\author[0000-0001-9158-0829]{Nicolas Lehner}
\affil{Department of Physics, University of Notre Dame, 225 Nieuwland Science Hall, Notre Dame, IN 46556}

\author[0000-0002-0355-0134]{Jessica K. Werk}
\affil{Department of Astronomy, University of Washington, Box 351580, Seattle, WA 98195}

\author[0000-0002-3120-7173]{Rongmon Bordoloi}
\affil{Department of Physics, North Carolina State University, 2401 Stinson Drive, Raleigh, NC 27695}

\author[0000-0003-1455-8788]{Molly S. Peeples}
\affil{Space Telescope Science Institute, 3700 San Martin Drive, Baltimore, MD 21218}
\affil{Department of Physics and Astronomy, Johns Hopkins University, 3400 N. Charles Street, Baltimore, MD 21218}



\begin{abstract}
We present new calculations of the mass inflow and outflow rates around the Milky Way (MW), derived
from a catalog of ultraviolet metal-line high velocity clouds (HVCs).
These calculations are conducted by 
transforming the HVC velocities into the Galactic Standard of Rest (GSR) reference frame, 
identifying inflowing ($v_{\rm GSR}\!<\!0$\kms) and outflowing ($v_{\rm GSR}\!>\!0$\kms) populations, and
using observational constraints on the distance, metallicity, dust content, covering fractions, 
and total silicon column density of each population. 
After removing HVCs associated with the Magellanic Stream and the Fermi Bubbles, we find 
inflow and outflow rates in cool ($T\!\sim\!10^4$\,K) ionized gas of
$dM_{\rm in}/dt\!\ga\!(0.53\pm0.23)\,(d/12\,{\rm kpc})(Z/0.2Z_\odot)^{-1}$\smy\ and 
$dM_{\rm out}/dt\!\ga\!(0.16\pm0.07)\,(d/12\,{\rm kpc})(Z/0.5Z_\odot)^{-1}$\smy. 
The apparent excess of inflowing over outflowing gas suggests that the MW is currently 
in an inflow-dominated phase, but the presence of substantial mass flux in both directions 
supports a Galactic fountain model, in which gas is constantly recycled between the disk and the halo. 
We also find that the \emph{metal flux} in both directions (in and out) is  indistinguishable.
By comparing the outflow rate to the Galactic star formation rate, we present the first estimate of 
the mass loading factor ($\eta_{\rm HVC}$) of the disk-wide MW wind, finding 
$\eta_{\rm HVC}\ga(0.10\!\pm\!0.06)\,(d/12\,{\rm kpc})(Z/0.5Z_\odot)^{-1}$. 
Including the contributions from low- and intermediate-velocity clouds and from hot gas
would increase these inflow and outflow estimates.
\end{abstract}

\section{Introduction} \label{sec:intro}
It has long been known that the Milky Way (MW) accretes gas from its surroundings \citep{Oo69},
allowing it to sustain its star formation over long ($\sim$Gyr) timescales \citep{La80}. 
In turn, this stellar activity drives gas back into the halo of the Galaxy 
through stellar winds and supernova feedback. Star formation is thus regulated by a
symbiotic relationship between inflow and outflow, in which inter-dependent gas flows play
an important role in the baryon cycle \citep{Pu12, Tu17}.
Observationally characterizing the Galactic
mass inflow and outflow rates is of high importance for validating this picture.
Studying gas flows around the MW (as opposed to external galaxies) 
offers several key advantages include proximity, abundance of data,
and knowledge of nearby Galactic structure.

Ultraviolet (UV) absorption-line studies are particularly powerful for 
inflow and outflow studies 
due to their sensitivity and ability to trace multiple ISM gas phases.
In order to identify inflowing and outflowing UV absorbers, foreground
components associated with the Galactic ISM need to be removed. 
Traditionally this is achieved by classifying interstellar UV absorbers 
into three populations using their velocity relative to the 
Local Standard of Rest (LSR): 
low-velocity clouds (LVCs, $|v_{\rm LSR}|<40$\kms),
intermediate-velocity clouds (IVCs, $40<|v_{\rm LSR}|<90$\kms), and
high-velocity clouds (HVCs, $|v_{\rm LSR}|>90$\kms).
HVCs are moving too fast for co-rotation with the Galactic disk, and 
instead represent inflowing or outflowing gas. As such, they are 
direct tracers of accretion and feedback and have well-characterized 
neutral and ionized gas properties
\citep{WW97, Se03, Sh09, Co09, LH11, Le12, He13, Fo14, R17chap}. 

In this paper, we present new calculations of the mass inflow 
and outflow rate represented by HVCs, using a new 
kinematic distinction into inflowing and outflowing populations.
This allows us to trace the circulation of gas and metals
and hence to derive the \emph{net} inflow rate, the metal flow rates,
and the mass loading factor of the outflowing gas, all of 
which have not been derived before. The paper is structured as follows.
In Section~\ref{sec:calc} we present the equations used to derive the HVC masses and mass flow rates. In Section~\ref{sec:values} we describe the adopted values of HVC 
properties used to evaluate the flow-rate equations, and quantify their uncertainties.
In Section~\ref{sec:discussion} we discuss our results and their astrophysical implications. 
Appendix~\ref{sec:ea} describes the propagation of errors though our analysis.

\section{Derivation of HVC Mass and Mass Flow Rate} \label{sec:calc}

Following earlier work \citep[][R17]{Wa08, Sh09, LH11, Le12, Fo14}, 
we derive the gas mass and mass flow rates of HVCs 
from UV metal-line absorption measurements. The principal difference 
between our new calculations and earlier work is that we split the 
HVC population into inflowing and outflowing components, and conduct 
the analysis separately for each subset. 

\begin{figure*}[!ht]
    \centering
    \includegraphics[width=\textwidth]{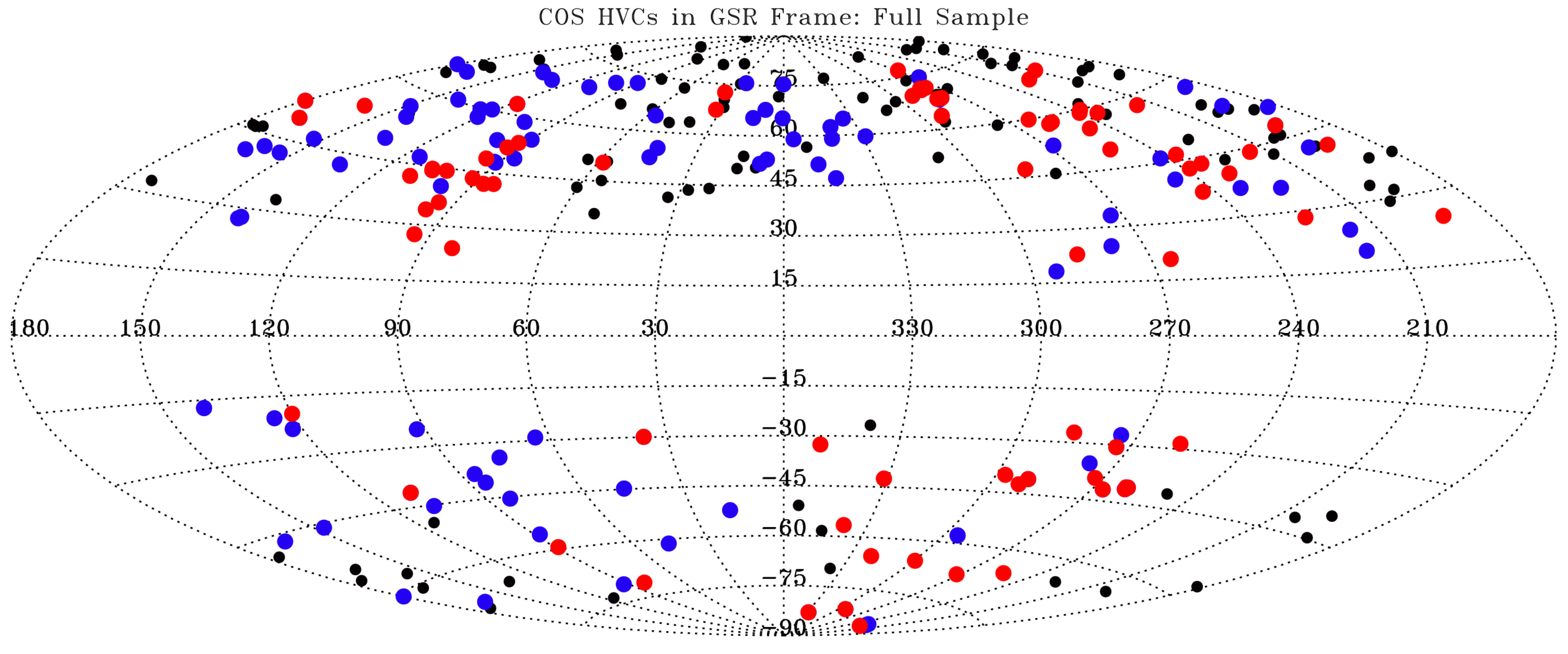}
    \includegraphics[width=\textwidth]{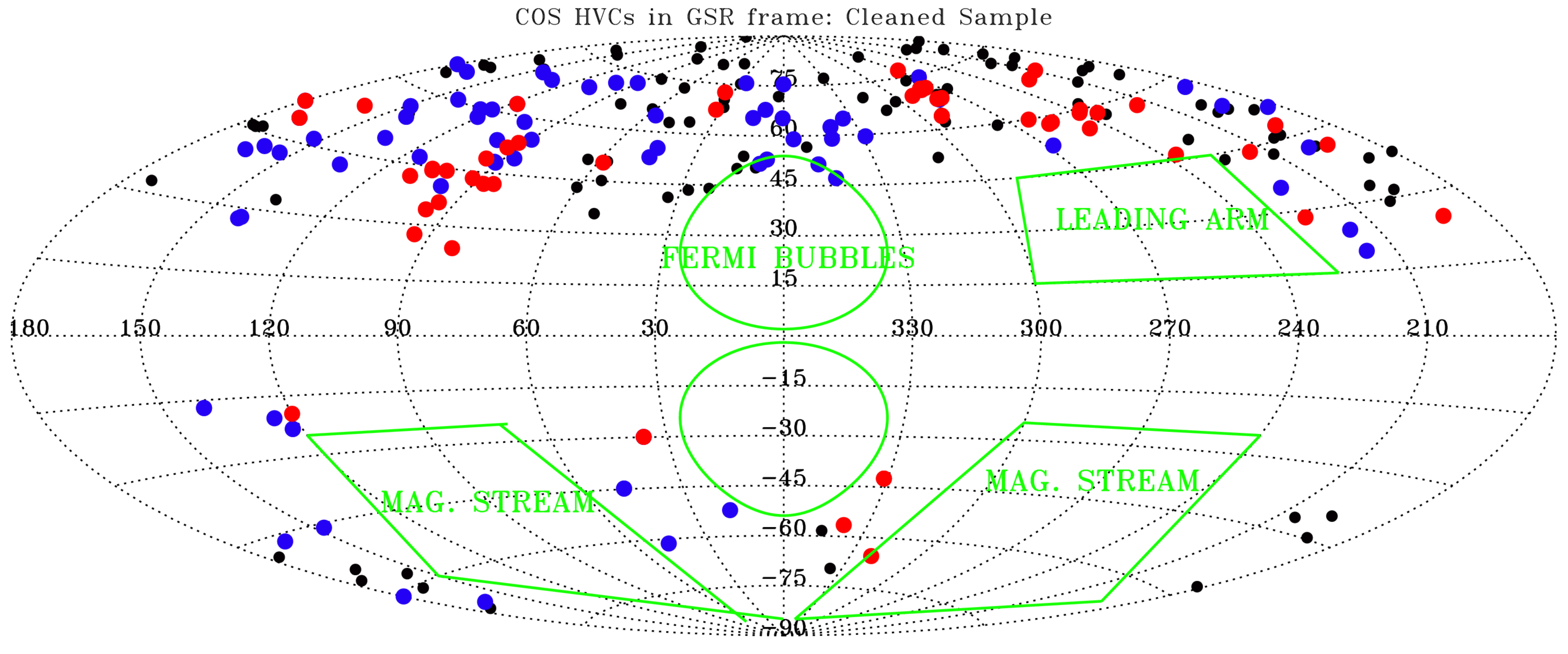}
    \caption{{\bf Top:} All-sky map in galactic coordinates of the HVCs in the R17 sample, color-coded by their 
    GSR velocity. Inflowing ($v_{\rm GSR}\!<\!0$\kms) HVCs are shown in blue, 
    outflowing ($v_{\rm GSR}\!>\!0$\kms) HVCs are shown in red, and sightlines with no HVCs detections
    are shown in black. 
    {\bf Bottom:} Same as the top but showing the cleaned sample, i.e. with sightlines passing through the 
    Magellanic Stream, Leading Arm, and Fermi Bubbles removed. The outline of these structures is shown in green.
    Cleaning the sample reduces the sample size, but only slightly reduces the covering fractions.}
    \label{fig:map1}
\end{figure*}

We begin with the R17 HVC catalog of 187 metal-line HVCs detected in 
270 extragalactic sightlines surveyed with the Cosmic Origins Spectrograph 
\citep[COS;][]{Gr12} on the \emph{Hubble Space Telescope} (\emph{HST}). 
All 187 HVCs are securely detected, meaning they are present at 
$>$4$\sigma$ significance in at least two metal-line transitions
(typically \siw\ $\lambda$1260,1193,1190,1304, \sit\ $\lambda$1206, or \cw\ $\lambda$1334).
The sightlines are spread fairly evenly across the Northern and Southern Galactic 
hemispheres at $|b|\!>\!15\degr$, and also show an approximate East-West symmetry
(see Figure~\ref{fig:map1}).
By definition, the HVCs are selected in the LSR reference frame,
all having either $v_{\rm LSR}\!<\!-90$\kms\ or $v_{\rm LSR}\!>\!90$\kms.
R17 assessed the presence of sample biases, and found that while some
S/N-related and target-selection biases inevitably exist, these are likely
to be minor compared to the other systematics, 
such as the velocity cutoff used to select HVCs.

First we transform the HVC velocities from the LSR reference frame to 
the Galactic Standard of Rest (GSR), to remove the effect of Galactic rotation.

\begin{equation}
    v_{\rm GSR}=v_{\rm LSR}+240\,{\rm sin}\,l\,{\rm cos}\,b\,{\rm km\,s^{-1}},
    \label{eqn:gsr}
\end{equation}

where $l$ and $b$ are the Galactic longitude and latitude of the sightline,
and we adopt the rotation velocity at the solar circle of 240$\pm$8\kms\ 
from \citet{Re14}, based in interferometric measurements.
We show the sky distribution of the HVC population in the GSR reference frame in 
the top panel of Figure~\ref{fig:map1}, color-coded into 
inflowing ($v_{\rm GSR}\!<\!0$\kms) and outflowing 
($v_{\rm GSR}\!>\!0$\kms) components.
Note that while there is structure in the gas distribution, inflowing and outflowing 
clouds are often found close together in the same regions of the sky. 
    
Next, we produce a cleaned sample by removing HVCs that are associated 
with the Magellanic Stream and its Leading Arm, since these structures
are much more distant than the general HVC population \citep{Br05, Ni08, DO16}.
This is achieved by cross-matching the R17 sample with the \citet{Fo14} catalog of Stream 
absorbers, which covers sightlines within 30$\degr$ of the 21\,cm emission from the Stream,
and then removing matched sightlines from the sample.
We also remove several sightlines passing through the Fermi Bubbles \citep{Bo17, Ka18}, 
since the HVCs in these directions are thought to trace a
nuclear wind and so are distinct from the general HVC population. 

We then calculate the total (neutral plus ionized) hydrogen column density 
$N_{\rm H}$ in cool gas ($T\!\approx\!10^4$\,K) in each HVC using the silicon lines as a proxy, by summing 
over the observed ionization states of Si and then correcting for metallicity and dust.

\begin{equation}
    N_{\rm H}=N({\rm H\,I})+N({\rm H\,II})=\frac{N(\rm{Si\,II})+N(\rm{Si\,III})}{{\rm (Si/H)(10^{[Si/Zn]})}},
    \label{eqn:nh}
\end{equation}

where (Si/H) is the silicon abundance and the [Si/Zn] term quantifies 
the level of dust depletion, since Si depletes onto dust grains but 
Zn is largely undepleted \citep{Je09, Tc15}.
The ionic column densities are apparent optical depth measurements
of \siw\ (based on $\lambda$1260, 1190, 1193, 1304, 1526) and 
\sit\ $\lambda$1206 (see Table A.4 in R17). 
Note that this method does not account for metal atoms locked in unobserved
ionization states; it is based solely on the observed column
densities of the cool-gas tracers \siw\ and \sit. 

Our calculations adopt a two-component spherical-shell model,
in which inflowing and outflowing HVCs exist at a uniform distance from the disk
with different covering fractions
\citep[corrections can be made for ellipsoidal or cylindrical geometries, 
though these are minor;][]{Ri17}.
In this model the total gas mass 
contained in HVCs, $M$, is formed by combining the mean total hydrogen column 
$\langle N_{\rm H}\rangle$ with the measured HVC sky covering fraction 
$f_{\rm sky}$ and mean HVC distance $\langle d\rangle$. This calculation 
can be done separately for the inflowing and outflowing populations.

\begin{eqnarray}
    M_{\rm in} & = & 1.4m_{\rm H}f_{\rm sky, in} \langle N_{\rm H, in} \rangle\,4\pi d^2 \hspace{0.25cm}{\rm and} \label{eqn:mass}\\
    M_{\rm out}& = & 1.4m_{\rm H}f_{\rm sky, out}\langle N_{\rm H, out}\rangle\,4\pi d^2,
    \label{eqn:mass2}
\end{eqnarray}

where $m_{\rm H}$ is the mass of the hydrogen atom, the factor 1.4 
accounts for the mass in helium and metals, and the covering fractions
are assumed to be independent of $d$.
Each of the terms on the right hand side of Equations~\ref{eqn:mass} and 
\ref{eqn:mass2} can be directly constrained by observations.

Finally, we combine the gas mass estimates with the mean flow velocities 
$\langle v_{\rm in}\rangle$ and $\langle v_{\rm out}\rangle$
to derive the mass flow rates $dM/dt$ for the inflowing and 
outflowing HVC populations.

\begin{eqnarray}
    dM_{\rm in}/dt & = & M_{\rm in}\langle v_{\rm in} \rangle/d \hspace{0.25cm}{\rm and}
    \label{eqn:massflow}\\
    dM_{\rm out}/dt& = & M_{\rm out}\langle v_{\rm out} \rangle/d.
    \label{eqn:massflow2}
\end{eqnarray}

The difference between the inflow rate and the outflow rate
gives the \emph{net} inflow rate, and the ratio of the mass outflow 
rate to the star formation rate (SFR) gives the HVC mass loading 
factor, $\eta_{\rm HVC}$, which quantifies the effectiveness of stellar 
feedback in driving and accelerating the outflow. 

\begin{equation}
    \eta_{\rm HVC}=\frac{dM_{\rm out}/dt}{\rm SFR}.
    \label{eqn:eta}
\end{equation}

\begin{deluxetable*}{lll ll}[!ht]
\tabcolsep=4.0pt
\tablewidth{0pt}
\tablecaption{Covering Fractions of Inflowing and Outflowing HVCs}
\tablehead{Population & 
$f_{\rm sky, raw}$\tm{a} & $f_{\rm sky, clean}$\tm{a} &$\langle v_{\rm GSR}\rangle$\tm{b} & $\langle$log\,$N_{\rm H}\,\rangle$\tm{c}  \\
& & & (\kms) & ($N$ in cm$^{-2}$) }
\startdata
Inflowing  & 0.39$\pm$0.03 (106/270) & 0.36$\pm$0.03 (75/211) & $-$101$\pm$10 & 18.93$\pm$0.06\\
Outflowing & 0.30$\pm$0.03 (81/270)  & 0.25$\pm$0.03 (52/211) &    +89$\pm$12 & 18.62$\pm$0.08\\
\enddata 
\tn{a}{Sky covering fraction of the raw (R17) and cleaned (non-Magellanic, non-Fermi Bubble) HVC sample. 
Errors are based on the Wilson score interval.}
\tn{b}{Mean GSR velocity with its standard error.} 
\tn{c}{Mean total hydrogen column density from Equation~\ref{eqn:nh} with its 
standard error, assuming a metallicity of 20\% solar for the inflowing gas 
and 50\% solar for the outflowing gas.
Value does not account for potential saturation of \sit\ $\lambda$1206.}
\label{tab:inputs}
\end{deluxetable*}

\begin{deluxetable*}{lllll ll}[!ht]
\tabcolsep=4.0pt
\tablewidth{0pt}
\tablecaption{HVC Masses, Mass Flow Rates, and Mass Loading Factors}
\tablehead{Distance & $M_{\rm in}$ & $M_{\rm out}$ & $dM_{\rm in}/dt$ & $dM_{\rm out}/dt$ & $dM_{\rm net}/dt$ & $\eta_{\rm HVC}$\tm{a}\\
(kpc) & (10$^7$ M$_\odot$) & (10$^7$ M$_\odot$) & (M$_\odot$\,yr$^{-1}$) & (M$_\odot$\,yr$^{-1}$)  & (M$_\odot$\,yr$^{-1}$) & }
\startdata
5  & 1.1$\pm$0.5 & 0.4$\pm$0.2 & $-$0.22$\pm$0.09 & 0.07$\pm$0.03 & $-$0.15$\pm$0.09 & 0.04$\pm$0.02\\
10 & 4.2$\pm$1.7 & 1.4$\pm$0.6 & $-$0.44$\pm$0.19 & 0.13$\pm$0.06 & $-$0.31$\pm$0.20 & 0.08$\pm$0.05\\
12 & 6.1$\pm$2.5 & 2.1$\pm$0.9 & $-$0.53$\pm$0.23 & 0.16$\pm$0.07 & $-$0.37$\pm$0.24 & 0.10$\pm$0.06\\
15 & 9.6$\pm$4.0 & 3.3$\pm$1.5 & $-$0.66$\pm$0.28 & 0.20$\pm$0.09 & $-$0.50$\pm$0.29 & 0.12$\pm$0.07\\
\enddata 
\tablecomments{All values are calculated using Equations \ref{eqn:mass} to \ref{eqn:eta},
for four assumed distances, chosen to bracket observations, and using 20\% solar metallicity for the inflow
and 50\% solar metallicity for the outflow. Negative flow rates
denote infall, so the net flux is inward for all distance cases.}
\tn{a}{$\eta_{\rm HVC}$=($dM_{\rm out}/dt$)/SFR with an SFR of 1.65\smy\ \citep{LN15}.}
\label{tab:values}
\end{deluxetable*}

\section{Adopted Values for HVC Parameters} \label{sec:values}

In this section we describe and justify the choice of HVC parameters adopted 
in the mass flow rate calculations, and quantify their uncertainties.
The covering fractions for the inflowing and outflowing HVC populations
are given in Table~\ref{tab:inputs}, together with their mean GSR velocities and mean
total hydrogen column densities.
The resulting values for the HVC masses, mass flow rates, and mass loading factors (with errors)
calculated using Equations \ref{eqn:mass} to \ref{eqn:eta}
are given in Table~\ref{tab:values}. 

\subsection{Covering Fraction} \label{fcov}

The measured sky covering fractions for \emph{inflowing} HVCs are 
39$\pm$3\% (106/270) in the full sample and 36$\pm$3\% (75/211) in 
the cleaned sample. For \emph{outflowing} HVCs, the covering fractions are 
30$\pm$3\% (81/270) in the full sample and 25$\pm$3\% (52/211) in the cleaned sample.
The errors here reflect the Wilson score interval.
Although the cleaned sample is smaller in size, particularly in the South
where most Magellanic gas resides,
the covering fractions are only slightly reduced by the 
sample-cleaning process (see Figure~\ref{fig:map1}), because cleaning removes
some directions with non-detections as well as detections.
The total sky covering fraction $f_{\rm sky, total}=f_{\rm sky,in}+f_{\rm sky, out}$=69$\pm$4\%
for the full sample and 61$\pm$4\% for the cleaned sample,
in good agreement with values measured in previous UV HVC surveys:
\citet{Le12} report $\approx$68\%, 
\citet{Sh09} report 81$\pm$5\% for a \sit\ survey,
and \citet{Se03} report 58\% for an \ion{O}{6} survey. 
The \sit\ and \ion{O}{6} surveys have different values because they have differing sensitivities and only require a detection in a single absorption line, whereas 
R17 and \citep{Le12} require a detection in at least two lines.
For comparison, the global 21-cm HVC sky covering fraction is 37\% down to $N$(\hi)=$7\times10^{17}$\sqcm\ \citep{Mu95}.
The much higher covering fraction in the UV surveys is due
to their higher sensitivity: UV HVCs are sensitive to gas down to log\,$N$(\hi)$\sim$14 \citep{Fo06}.

The cleaned sample is largely (but not completely) contained in the northern 
hemisphere, because so much Magellanic gas is removed in the south.
If we restrict our analysis to the northern hemisphere only, we find a 
cleaned covering fraction of 33$\pm$3\% for inflowing HVCs and 
25$\pm$3\% for outflowing HVCs, very close to 
the values obtained from the combined (N+S) sample. 
Thus we conclude there is no evidence for bias 
from using a predominantly northern sample, and we proceed with the covering fractions from 
the combined (N+S) sample. The {\it non-detections} are distributed across the northern and southern 
hemispheres, and across the east and west. However, there is some structure in their distribution,
with clusters of non-detections found in three regions (see Figure~\ref{fig:map1}): one in the far north
at $|b|\!>\!75$\degr, one in the north at $30\!<\!b\!<\!45$\degr, 
$0\!<\!l\!<\!60$\degr\ (east of the Fermi Bubble), and one in the south
at $30\!<\!b\!<\!45$\degr, $0\!<\!l\!<\!60$\degr\ (west of the Magellanic Stream). 
These indicate that the HVC sky distribution
is patchy, as discussed in earlier surveys \citep[][R17]{Se03, Fo06, Le12}.

\subsection{Metallicity} \label{sec:metallicity}

We adopt a 20\% solar metallicity for the inflowing gas, 
(Si/H)$_{\rm in}$=0.2(Si/H)$_\odot$,
based on the observed metallicities of the infalling HVC Complex C
\citep{Wa99, Ri01a, Gi01, Co03, Tr03, Sh11}. 
\citet{Fo04} updated all the Complex C metallicity measurements 
onto a common solar abundance scale,
and found a mean value of 20$\pm$4\% solar, so we adopt a 20\% (4/20) error.
We adopt a 50\% solar metallicity for the outflowing gas, 
(Si/H)$_{\rm out}$=0.5(Si/H)$_\odot$, as seen in the Smith Cloud 
\citep{Fo16}, one of the only examples of a metal-enriched
(and therefore Galactic in origin) HVC \citep[see also][]{Ze08}.
We note there are fewer metallicity 
measurements for outflowing HVCs than for inflowing HVCs. 
Nonetheless, expectations from theory are that outflowing gas should be
metal-enriched, particularly if driven out by feedback from star formation.
Furthermore, IVCs are observed to show $\approx$0.5--1 solar metallicities, 
indicating that metal-enriched gas exists at lower velocities closer to the disk
\citep{Wa01, Ri01a, R17chap}.

\subsection{Dust depletion} \label{sec:dust}

For the dust depletion, we adopt an average Si depletion [Si/Zn]=$-$0.26$\pm$0.14 
measured in MW halo clouds from \citet{SS96}, taken to be applicable to HVCs.
This has the effect of increasing the total HVC mass by a factor of 1.8.
The presence of dust in HVCs is supported by 
the detection of far-IR emission from Complex C \citep{MD05},
the detection of molecular hydrogen in the Magellanic Stream  
\citep{Ri01b}, since H$_2$ forms on the surface of dust grains, and by
UV depletion measurements of the Stream's chemical composition \citep{Fo13}.
However, IR emission studies of other HVCs found no evidence 
for dust \citep{Wi12, Sa14}, so the average amount of dust in the HVC population
is unknown, and could be small. As such, the dust correction factor of 1.8 is uncertain.
We estimate an error of 38\% using the dispersion in [Si/Zn] in halo clouds \citep{SS96}.

\subsection{Distances} \label{sec:distances}

For the HVC distances, we consider four values: 5, 10, 12, and 15 kpc.
These values bracket the observed values of both \emph{statistical} HVC distances 
\citep{LH11}, which are derived from the similar 
incidence rates of HVCs toward stars and toward AGN, 
and \emph{individual} HVC distances 
\citep{Wa01, Wa08, Th08, LH10, Sm11, Ri15, Pe16}, 
which are based on the detection of HVCs in absorption toward stellar 
targets at known distance. Both methods find HVC distances in 
the range $\approx$5--15 kpc (not including the Magellanic Stream).
\citet{LH11} report a mean value $\langle d\rangle$=12$\pm$4\,kpc, 
so we adopt this as our nominal distance with a 33\% uncertainty (=4/12).

\subsection{Galactic Star Formation Rate} \label{sec:sfr}

For the Galactic SFR, published measurements range from
0.68--1.45\smy\ \citep{RW10}, 1.65$\pm$0.19\smy\ \citep{LN15},
1.9$\pm$0.4\smy\ \citep{CP11}, to 3.8$\pm$2.2\smy\ \citep{Di06}, 
where the differences in part reflect the choice of initial mass function adopted. 
For example, changing a Kroupa IMF to a Salpeter IMF
would increase the derived SFR by a factor of 1.4.
We adopt the recent \citet{LN15} value, though with a larger uncertainty 
to reflect the 40\% IMF error, i.e. 1.7$\pm$0.7\smy. This propagates to the error
on the mass loading factor.

\subsection{Velocity Limit, Ionization, and Saturation Effects}

For several reasons, the HVC inflow and outflow rates calculated using 
Equations~\ref{eqn:gsr} to \ref{eqn:massflow2} are lower limits on the true 
inflow and outflow rates. 
First, much of the MW halo gas is 
unobservable in UV absorption because any gas moving at low velocities 
blends with the strong foreground ISM absorption.
\citet{Zh15} used simulations to show that approximately half of the halo gas mass 
is unobservable because of this effect \citep[see also][]{Zh19, QB19}, so 
we should keep in mind that our calculations only refer to gas at HVC velocities.
Second, our total hydrogen column density calculations have not accounted for 
mass in \ion{Si}{4} or unobserved higher ionization states of silicon. 
We do not include \sif\ because the metallicity in the high-ion phase is unknown,
and could be different than in the low-ion phase.
Any hot ($T\!>\!10^6$\,K) outflowing gas is not detectable in UV observations,
but according to simulations could bear a substantial fraction of the mass budget 
\citep[e.g.][]{Fi18, Ki18}.
Third, some unresolved saturation may exist in \sit\
$\lambda$1206, a strong line that 
is often observed to be saturated in HVCs \citep[][R17]{Sh09, Co09}. 
Above log\,$N$(\sit)=12.53, the optical depth at line center becomes $>$1 for a
line width $b$=10\kms, and so saturation becomes potentially important.
\sit\ contributes a large fraction of the total silicon mass since it is 
often the dominant ionization stage of Si, so saturation of \sit\ will affect the total Si column.
Accounting for each of these effects would serve to
increase $\langle N_{\rm H}\rangle$ and therefore the mass flow rates.
Finally, the HVC mass flow rates presented here do not include (by design) the 
Magellanic Stream and Leading Arm or the Fermi Bubbles, since these structures are 
physically unrelated to the remaining HVC population. The Stream represents an
a much larger inflow rate of $\approx$3--7\smy\ \citep[][R17]{Fo14}
and the Fermi Bubbles have an outflow rate of 0.2--0.3\smy\ in cool gas \citep{Bo17}.


\section{Discussion and Summary} \label{sec:discussion}

We have derived new empirical constraints on the mass inflow and outflow 
rates of cool ionized gas around the Milky Way, by dividing the R17 catalog of 187 
UV absorption-line HVCs into inflowing and outflowing populations,
and using direct observational constraints 
on HVC metallicity, dust content, and distance. 
After removing HVCs associated with the Magellanic Stream and Fermi Bubbles, we find 
$dM_{\rm in}/dt\ga(0.53\pm0.23)\,(d/12$\,kpc)$(Z/0.2Z_\odot)^{-1}$\smy\ and 
$dM_{\rm out}/dt\ga(0.16\pm0.07)\,(d/12$\,kpc)$(Z/0.5Z_\odot)^{-1}$\smy, 
where we have scaled the results to 20\% solar metallicity for the inflow
and 50\% solar metallicity for the outflow, as HVC observations indicate,
and where the uncertainties are formed by propagating the errors discussed in 
Section~\ref{sec:values} on $f_{\rm sky}$, $\langle N_{\rm H}\rangle$, 
$\langle v_{\rm GSR} \rangle$, and dust depletion 
(full details on the error propagation are given in Appendix~\ref{sec:ea}.)
If we also propagate the errors on metallicity and distance instead of 
leaving them as free parameters, we derive final values of
$dM_{\rm in}/dt\ga(0.53\!\pm\!0.31)$\smy\ and 
$dM_{\rm out}/dt\ga(0.16\!\pm\!0.10)$\smy, 
where the ``$\ga$'' symbol remains because of potential line saturation in \sit\ 1206.

A comparison of these flow rates shows that:
(1) the Milky Way appears to be currently in an inflow-dominated phase, 
since we observe a net inflow \emph{regardless of HVC distance},
although this is subject to large accumulated uncertainties.
Nonetheless, the outflow metallicity would need to be 0.15$Z_\odot$ 
for the inflow and outflow rates to match, and 
such a low value is highly unlikely for a Galactic wind.
Furthermore, the excess of inflow over outflow is seen at high significance
in the sky covering fractions alone (Table~\ref{tab:inputs}.
(2) The HVC inflow rate is not sufficient to sustain the Galactic
SFR of 1.7$\pm$0.7\smy\ \citep{LN15}
unless hidden saturation of \sit\ 1206 causes our inflow rate to be underestimated. 
If saturation is not the dominant uncertainty, then the inflow metallicity 
would need to be 0.06$Z_\odot$ to match the SFR, and such low
metallicities are not measured anywhere in the MW gaseous halo. 
(3) the difference between the mass inflow and outflow rates is
largely driven by the difference in metallicity.
Put differently, the \emph{metal} inflow and outflow rates in HVCs, 
$dM_{Z, {\rm in}}/dt$=($Z$/H)$_{\rm in}dM_{\rm in}/dt$ and 
$dM_{Z, {\rm out}}/dt$=($Z$/H)$_{\rm out}dM_{\rm out}/dt$,
are statistically indistinguishable: we calculate 
$dM_{Z, {\rm in}}/dt\approx$18$\pm$8 milli-\smy\,($d$/12\,kpc) whereas 
$dM_{Z, {\rm out}}/dt\approx$14$\pm$6 milli-\smy\,($d$/12\,kpc). 
These metal fluxes are better known than the mass fluxes since 
they are independent of metallicity, and simply depend on summing 
over the observed ionization states of silicon and correcting for dust.

Despite the global excess of inflow, the presence of substantial gas flow in both directions 
lends support to the presence of a Galactic fountain 
\citep{SF76, Br80, HB90, Sp08, Ki18}, 
in which supernova-driven outflows drive metal-enriched gas 
$\sim$1--4\,kpc into the halo before it cools and cycle 
back to the plane on $\sim$50\,Myr timescales. 
Other recent observational studies of MW halo absorbers have also
found evidence for a fountain scenario \citep{Ma17, We19}.
Furthermore, our inflow rate of $\ga$0.53$\pm$0.31\smy\ matches the rate 
needed by Galactic Chemical Evolution models \citep{Ch01, Ch09} 
to explain stellar metal abundance patterns and to resolve the G-dwarf problem,
although we caution that the accumulated error is substantial. 
This confirms the finding of previous UV surveys that inflowing HVCs
trace the replenishment of gas for star formation \citep[][R17]{Sh09, LH11}.

Our inflow rate of $\ga$0.53$\pm$0.31\smy\ 
is of similar order to earlier UV-based measurements \citep[$\sim$1\smy;][R17]{Sh09, LH11},
but higher than 21-cm-based measurements derived from 
large HVCs \citep[$\approx$0.08\smy;][]{Pu12}
and higher than the flow rates derived for Complex C
\citep[$\approx$0.1\smy;][]{Wa99, Co03, Sh11}.
This makes sense since most HVC mass is in ionized gas \citep{LH11}.
The ratio of the UV inflow rate to the 21-cm inflow rate quantifies
the relative mass in ionized and neutral gas. 
Our inflow rate calculations indicate $\langle N$(\hw)/$N$(\hi)$\rangle\approx0.53/0.08=6.6$, 
equivalent to a mean HVC ionization fraction 
$f_{\rm H\,II}$=$N$(\hw)/$N$(\hi+\hw)=0.87. 
This is higher than the mean ionization fraction reported in large 21-cm HVCs such as 
Complex C and Complex A because such large clouds are atypical; 
most HVCs are ionized and not detected at 21\,cm \citep[][R17]{Sh09, LH11}.

For comparison with the HVC mass flow rates, the {\it intermediate} velocity 
mass flow rates are of interest. These can be derived from optical observations 
of \ion{Ca}{2} and \ion{Na}{1} absorption
\citep{BB12}. The \ion{Ca}{2} and \ion{Na}{1} doublets are much less 
saturated than UV resonance lines, particularly at intermediate velocities,
resulting in cleaner measurements.
\citet{BB12} report a population of \ion{Ca}{2} IVCs with 
a sky covering factor of $\approx$30\%, a mean $v_{\rm GSR}$=$-$42\kms, 
and an associated $N$(\hi) distribution 
peaking near log\,$N$(\hi)$\approx$19. Combining these numbers with a typical 
IVC distance of 2\,kpc \citep{Ri17} we derive an
\hi\ mass flow rate of $\approx$0.04\smy\ in these \ion{Ca}{2} IVCs.
If a similar ionization correction $N$(\hw)/$N$(\hi)=6.6 
applies to IVCs as to HVCs, then IVCs represent a total mass 
flow rate of $\approx$0.27\smy, comparable to the HVC flow rate
(this similarity occurs because the larger column densities of IVCs are compensated 
by their smaller velocities and distances). To fully explore this, 
a more thorough investigation of the IVC mass flow rates using UV data would be worthwhile,
including a distinction into inflow and outflow and detailed ionization modeling. 

We emphasize that the HVC mass flow rates derived in this paper are 
effectively \emph{instantaneous}, since they represent the mass 
flux contained in HVCs observed at the current epoch. 
Hydrodynamic simulations \citep{HP09, Ar17} show that HVC lifetimes are typically 
a few 100\,Myr depending on HVC mass and the density contrast with 
the surrounding hot medium, after which the clouds will become disrupted and
evaporated by coronal interactions. Caution should therefore be 
used in integrating our mass flow rates over long timescales to calculate 
the total mass delivered in and out of the disk, since the gas will change 
phase over time via evaporation and subsequent condensation 
\citep[e.g.][]{Fr17, Vo19}.

Finally, we can use our HVC mass outflow rate measurement to present
the first observational derivation of the mass loading factor 
of the disk-wide MW wind. We derive
$\eta_{\rm HVC}\!\ga\!(0.10\pm0.06)\,(d/12\,{\rm kpc})(Z/0.5Z_\odot)^{-1}$. 
This matches the predictions from the numerical simulations of feedback by 
\citet{Ki18}, who find $\eta$=0.1, 
but is lower than expectations from chemical evolution modeling;
for a MW-type galaxy at $z$=0, \citet{PS11} derive $\eta\approx1.4$ while 
\citet{BB18} derive $\eta\approx2$.
However, our value for $\eta_{\rm HVC}$
is a lower limit for the same reasons that the mass
outflow rate is a lower limit: saturation, ionization effects, and velocity effects. 
Furthermore, it does not include the contribution 
of the nuclear wind at the Galactic Center. 
According to biconical outflow modeling of UV absorbers in  
Fermi Bubble sightlines, the nuclear wind represents a cool gas outflow 
rate of 0.2--0.3\smy\ \citep{Fo15, Bo17} and an associated mass loading factor
$\eta_{\rm nuclear}$=0.12--0.18. The similarity of the nuclear and disk-wide
outflow rates (they agree to within a factor of two) indicates that the Galaxy
releases metal-enriched gas into the halo
through nuclear feedback \emph{and} disk-wide feedback, 
and both should be considered active channels 
through which it exchanges mass with its environment.

\vspace{0.5cm}
{\it Acknowledgments.}
We are grateful to the referee for a constructive report that improved the manuscript.
Support for program 15020 was provided by NASA through grants from the Space 
Telescope Science Institute, which is operated by the Association of Universities 
for Research in Astronomy, Inc., under NASA contract NAS~5-26555.
We acknowledge useful conversations on mass flow rates with Josh Peek.\\

{\it Facilities:} HST (COS)\\

\appendix

\section{Error Analysis} \label{sec:ea}

In Table~\ref{tab:errors} we document the error analysis used to derive the uncertainties on the HVC masses, mass flow rates, metal flow rates, and mass loading factors. The rows in the table show how the percentage errors on each measured quantity (sky covering fraction, mean total silicon column, dust depletion, mean GSR velocity, SFR, metallicity, and distance) propagate to the errors on the derived quantities. The total percentage errors, $\sigma$(total), are calculated by adding the input percentage errors in quadrature since $M$, $dM/dt$, $dM_Z/dt$, and $\eta_{\rm HVC}$ are all formed via products and division (Equations~\ref{eqn:mass} to \ref{eqn:eta}). 
The dominant error terms are those on the dust, metallicity, and distance, and also SFR, though that only impacts the mass loading factors. The error terms on the covering fractions, mean total silicon column, and mean GSR velocity are much smaller since those quantities are better characterized observationally.
We also compute a second total error $\sigma$(reduced), which is the same as $\sigma$(total) but without the contribution
from the distance and metallicity errors. This is because in the text and in Table~\ref{tab:values} we 
leave distance and metallicity as scalable parameters, so that the equations can be re-evaluated if better constraints become available 
in the future.

\begin{deluxetable*}{l lllllll ll}
\tabcolsep=5.0pt
\tablewidth{0pt}
\tablecaption{Propagation of Errors}
\tablehead{Quantity & $\sigma(f_{\rm sky})$ & $\sigma(\langle N_{\rm Si}\rangle)$ & 
$\sigma$([Si/Zn]) & $\sigma(\langle v_{\rm GSR}\rangle)$ & $\sigma$(SFR) &
$\sigma$([Si/H]) & $\sigma(d)$ & $\sigma$(total) & $\sigma$(reduced)\\
 & (\%)& (\%)& (\%)& (\%)& (\%)& (\%)& (\%)& (\%) & (\%)}
\startdata
$M_{\rm in}$           &  8 & 15 & 38 & ... & ... & 20 & 33 & 66 & 42\\
$M_{\rm out}$          & 12 & 20 & 38 & ... & ... & 20 & 33 & 68 & 45\\
$dM_{\rm in}/dt$       &  8 & 15 & 38 &  10 & ... & 20 & 33 & 58 & 43\\
$dM_{\rm out}/dt$      & 12 & 20 & 38 &  13 & ... & 20 & 33 & 60 & 46\\
$dM_{Z, {\rm in}}/dt$  &  8 & 15 & 38 &  10 & ... & ...& 33 & 54 & 43\\
$dM_{Z, {\rm out}}/dt$ & 12 & 20 & 38 &  13 & ... & ...& 33 & 57 & 46\\
$\eta_{\rm HVC}$       & 12 & 20 & 38 &  13 &  40 & 20 & 33 & 72 & 61\\
\enddata 
\tablecomments{This table reports the percentage errors on the {\it measured} HVC quantities (sky covering fraction, mean total silicon column, dust depletion, mean GSR velocity, SFR, metallicity, and distance) and how they propagate to the {\it derived} HVC quantities (masses, mass flow rates, metal flow rates, and mass loading factors). The errors are summed in quadrature following Equations~\ref{eqn:mass} to \ref{eqn:eta} to form the total error $\sigma$(total). A second total error $\sigma$(reduced) is computed {\it without} the distance and metallicity uncertainties, for use when those two variables are left as scalable parameters. No entry (...) means the variable is not used.}
\label{tab:errors}
\end{deluxetable*}

\end{document}